# Composition based machine learning to predict phases & strength of refractory high entropy alloys

M. Sreenidhi Iyengar[1], M. K Anirudh[2], P.H. Anantha Desik[3], M.P. Phaniraj[4]*

[1] nSpire AI, San Francisco, CA, USA.

[2]EY, International Tech Park, Madhapur, Hyderabad, Telangana 500081

[3]Tata Consultancy Services, Hyderabad, India 500081

[4]Department of Metallurgical and Materials Engineering, Mahatma Gandhi Institute of Technology, Hyderabad, India 500075

*Corresponding author: mphaniraj@gmail.com

**Abstract**

Refractory high-entropy alloys can function at temperatures exceeding those of nickel-based superalloys. Aluminum, as an alloying element, contributes multiple advantageous characteristics to various high-temperature alloys. The Aluminum containing RHEAs have the potential of being the best high temperature materials. In the present study we use the machine learning(ML) technique to determine the phase and yield strength of aluminum containing RHEAs. In this regard, we created the Al-RHEA dataset from the published [1] compilation of RHEA data. We applied multiple ML algorithms to the training set and determined that the CatBoost algorithm gave the best performance. We optimized the hyperparameters of this algorithm and tested it for robustness using cross-validation methods. The CatBoost model predicts the yield strength of test data accurately ($R^2$=0.98). The algorithm was applied to estimate the yield strength for alloy compositions absent from our current dataset, achieving accurate predictions for these unrecorded alloys indicating that the model has learnt the underlying rules to predict the yield strength sufficiently. We then predict the effect of varying aluminum content on yield strength of RHEA. The model predictions were rationalized in view of published data on Al-RHEAs. We also developed the CatBoost classifier model that predicts the phases formed in the alloy of a given composition accurately. The cause for errors in phase prediction is discussed.

# Introduction

In the past two decades a significant share of materials research has focused on High Entropy Alloys (HEA), beginning with publications by Yeh [2] and Cantor [3]. As more information and knowledge was gained in the HEA field [4], it diversified into multicomponent, multiphase, low, medium, and high entropy alloys. The availability of experimental data on many alloy systems makes the application of Artificial Intelligence techniques such as machine learning (ML), a promising route for alloy design. ML models were used to predict modulus [5,6], hardness [7,8], strength [9], phases formed [10–12] and phase design [13] in high entropy alloys. A review of literature on ML models applied to HEAs up to the year 2021 can be found in Ref. [14].

Ni-based and Ni/Fe-based superalloys contain the alloying element aluminum, which forms aluminides ($Ni_3Al$) and thereby increases their strength and creep resistance. Aluminum enhances the alloy by forming a protective oxide layer that boosts both oxidation and corrosion resistance. Aluminum, by virtue of its low density, has the beneficial effect of reducing alloy density. HEAs made with refractory metals (RHEAs) have promising high temperature properties [15] and can compete with Ni-based superalloys [4]. Senkov et al. [16] reported that aluminum containing RHEAs exhibit enhanced strength at high temperatures compared to advanced nickel-based superalloys. It is important to explore the aluminum containing refractory metal alloys that give rise to exciting mechanical properties. Comprehensive insights into the mechanical properties of refractory high-entropy alloys have been detailed by Couziniè et al [1]. This data can be used to develop the ML model to explore the properties of refractory HEAs.

In the present study, several machine learning models were evaluated and the ML model with the best performance: CatBoost is used to predict yield strength of aluminum containing RHEAs.

The database for aluminum containing RHEA was created based on the compilation of published RHEA data [1] and used to develop the ML model. The model predicts the yield strength of alloys with new compositions that were not in the data set. There is no report, to the authors knowledge, on ML model to predict the yield strength of aluminum-containing refractory high-entropy alloys based on their compositional variations.

**2.0 Materials and Methods**

### 2.1 Database

The detailed dataset on mechanical properties of refractory 122 high-entropy alloys and other complex concentrated alloys, assembled by Couziniè et al. [1] encompasses publications spanning from 2010 to early 2018. This data included the alloy composition, types of microstructures, the metallurgical state, and the type of loading in which the properties such as yield strength were measured. For some compositions, the mechanical properties after different processing conditions were reported. The total data set, including the latter was 320 in number. The individual references from which the data was collected were referred to determine the experimental composition, type of microstructure and room temperature yield strength of the alloys. The data set of 320 RHEAs was collected in a Microsoft Excel sheet for analysis. In the data set 118 RHEAs had aluminum in the composition. The aluminum containing RHEA data set contained yield strength values at room temperature; high temperature data were not included in the data set. This data set of aluminum containing RHEA was used to develop regression and classification models.

The Al-RHEAs in the database contained 4-6 different alloying elements each. There was a total of 12 different alloying elements in the composition of RHEAs in the database. Figure 1 shows the range of elemental composition in the alloys used in the database. The elemental composition data for the alloys in our database vary, with a common range of 16.6 - 20.7 at. %, as depicted by the dashed lines in the accompanying figure. Five RHEAs contain less than 3 at. % cobalt.

### 2.2 Analyzing Machine Learning models

PyCaret [17], a Python-based Machine Learning library, facilitates an effective evaluation of several regression algorithms such as Linear, Lasso, Decision Tree, AdaBoost, Random Forest, and others, utilizing experimental data to measure performance by regression-specific metrics like R2, RMSE, and MAE.

$R^2$, or the coefficient of determination, is calculated by the ratio of the sum of squared residuals (SSR) to the total sum of squares (SST), indicating how much of the variance in the dependent variable is explained by the independent variables. It ranges from 0 to 1, with higher values indicating a better fit.

Root mean square error (RMSE) is derived as the square root of the sum of squared differences between predicted and actual values divided by the number of observations (N). A smaller RMSE suggests lower average prediction errors.

MAE, or mean absolute error, is assessed by averaging the absolute differences between the predicted outcomes and the actual data across all observations, reflecting the average error magnitude without direction bias.

All numerical computations were performed on a computer equipped with an Intel® i7 10th generation processor and an RTX 2060 GPU.

### 2.3 CatBoost Regressor & Classifier

Yandex introduced the CatBoost approach, representing the newest member of gradient boosting decision tree ML methods [18]. Distinguished by its utilization of symmetric trees, CatBoost stands out among other boosting algorithms, resulting in reduced prediction time. Additionally, CatBoost leverages ordered boosting. In this research, we applied the CatBoost framework (accessible at: https://github.com/catboost/catboost ) along with Python to analyze the yield strength in RHEAs.

We split the data into training (80%) and testing (20%) segments for developing the CatBoost regressor model. Model inputs included alloy compositions expressed in at. % of Al, Nb, Co, Hf, Ti, Zr, V, Mo, Ta, Si & W, and the processing conditions, and the yield strength was the output. Only those data that contained yield strength determined at room temperature were included in the dataset. The processing conditions were: Air cooled (AC), Hot isostatically pressed and annealed (HIP+A), and annealed (A). The CatBoost algorithm encodes categorical variables into numerical values 0 and 1 using the one hot encoding technique.

Optimal hyper-parameters for maximum depth and learning rate were determined by evaluating their impact on the root mean squared error. The learning rate was adjusted within the range of 0.01 to 0.5 for each maximum depth setting of 4, 6, and 8. The optimization process is halted if there is no change in RMSE after 100 iterations.

### 2.3.1 Cross-validation

K-fold validation [22] referenced widely in the literature, is employed as a robust method to prevent overfitting to training data in predictive modeling. This method partitions the initial training dataset into k distinct and exclusive subsets. During each iteration, one subset serves as the validation set, with the other k - 1 subsets used for training. The procedure is executed repeatedly across all subsets, averaging the model's performance metrics. In this study, 5-fold cross-validation was applied to our dataset to assess the regression model's efficacy.

A similar procedure was used to determine and develop the ML Classifier. The classifier model was built using the RHEA's composition and its processing conditions as the input variables. The data were split into training (75%) and test (25%) data sets. The output of the model consisted of the different types of phases that formed in the RHEA. The integrated function of the CatBoost library – MultiClass [23] was used to build the model because the number of classes (phases)  in the output is more than one. The optimized model was put to the cross-validation test to determine average accuracy of the classifier model. The confusion matrix was used to illustrate the model's accuracy.

## 3. Results and Discussion

### 3.1 Data analysis

In Figure 2a, the correlation coefficient between variables is depicted using a color map. The intensity of color saturation—brighter for higher positive correlations and darker for more pronounced negative correlations—visually distinguishes the strength of associations among the variables.

Figure 2b illustrates a correlation matrix depicting the relationship between yield strength and various input variables. The strong correlation of strength with temperature is common knowledge, however strength is a function of other variables too. This illustrates that basic correlation analysis alone doesn't elucidate the interplay among variables or their collective impact on strength. Hence, more advanced prediction models are required.

### 3.2 Evaluating the Efficacy of Machine Learning Regression Models

Table 1 presents the performance indicators for various machine learning models. Notably, the CatBoost regressor achieves the highest R² score of 0.847 and records the lowest values in RMSE and MAE, positioning it as the optimal model for the current study.

The evaluation metrics for each model were calculated using the standard hyperparameter settings provided by the PyCaret Machine Learning library throughout the training process. Enhancing the accuracy of the CatBoost model is possible by optimizing the hyperparameters specifically for the RHEA dataset.

### 3.3 Hyperparameter Optimization

Table 2 displays the minimum RMSE values recorded for various hyperparameter configurations, including learning rate (ranging from 0.01 to 0.9) and maximum depth (ranging from 4 to 8)—on both training and test datasets. The optimal configuration was found to be maximum depth of 6 and a learning rate of 0.1, which were applied in training the CatBoost model. Subsequently, the RMSE was monitored across iterations.

Figure 3a illustrates the decline in RMSE for the CatBoost model correlating with an increase in the number of iterations. Training concluded when there was no further reduction in RMSE value for 100 consecutive iterations.

Figure 3b demonstrates the model's accurate predictions for the test dataset (20% of the entire dataset), yielding an impressive $R^2$ value of 0.99.

### 3.4 Crossvalidation

Results from the 5-fold cross-validation of the CatBoost model are detailed in Table 3, with $R^2$ values achieved for each fold. The highest $R^2$ observed across the folds approached 0.99, while the average was ~0.95, suggesting both low bias and high accuracy in the model's performance. Figure 3b, illustrates the application of the Leave One Out Cross Validation (LOOCV) method, testing the trained model on previously unseen data. The results of these two tests show that the model has learnt sufficiently general rules to predict properties of the system.

### 3.5 Model predictions

*Effect of Aluminum content on yield strength*

In the database the aluminum content common to all alloys, except four alloys, is in the range: 16.6-20.7 at. % (see Figure 1). The model was employed to estimate the yield strength within this specified range for a given RHEA composition in the database. Figure 5 shows the predicted change in yield strength for the RHEA ($Al_xMo0.5NbTa0.5TiZr$) with increase in aluminum content. The experimental yield strength at 17.6% Al matches with predicted yield strength. However, the increase in aluminum content decreased the yield strength. This is in contradiction to traditional theory of solution strengthening. The latter would predict an increase in yield strength with increase of aluminum content. A similar decrease in experimentally determined yield strength upon increase in aluminum content was reported [25] in the RHEA: $TaNbVTiAl_{(x+0.2)}$. They suggested that the strengthening method was complex because of the covalent bond formed between aluminum and other atoms. The drop in yield strength with increase in aluminum content will be investigated experimentally by the authors in the near future.

## 3.5 Classifier model

A procedure similar to that for determining the regression model was followed to develop the classifier model. Table 5 shows the performance of various ML classifiers. The CatBoost classifier, which had the best performance and was chosen for hyperparameter optimization (Table 6). The optimized classifier was subjected to crossvalidation (Table 7). The average accuracy after crossvalidation is $R^2=0.976$. The optimized model was then tested using test data. Results from the test data are presented in Figure 6 through a confusion matrix, which indicates a strong alignment of the predictions with the experimental data across all compositions.

We now test the CatBoost classifier on new alloys, which were not in the database (Table 4). Figure 7 shows the predicted and measured outcomes on phases formed in the new alloys. The predictions for alloy composition with serial no. 1-6 and 9 in Table 4 agree with experimental observations. The predictions for alloy composition with serial no. 7,8 &10 are incorrect. These alloys were identified to contain BCC+BCC and BCC+B2 phases. Within the database of RHEA alloy compositions, there are 2 entries for BCC+BCC phases and 13 for BCC+B2 phases. The compositions that had BCC phase were all predicted accurately. The number of compositions that had BCC phase was 42. Thereby, the inaccuracy in prediction could be due to fewer data with similar phases in the database that was used to train the classifier.

## 4. Summary and Conclusions

We have developed a Machine Leaning based regression and classification model for aluminum containing Refractory High Entropy Alloys using composition and processing parameters as inputs. The CatBoost model exhibited superior performance across both regression and classification tasks. The regression model predicts the yield strength, and the classification model predicts the phases formed in the alloys. Both models' outputs were validated with published experimental data.

1. The model can predict the effect of a range of aluminum content on yield strength reasonably.
2. The model accurately forecasts the yield strength for new alloy compositions previously unrecorded in our database.
3. The classifier predicts the phases formed based on the input composition and processing conditions accurately.

**CRediT Authorship contribution statement**

**M.S.I**: Data collection, Formal analysis, Software, Methodology. **M.K.A**: Methodology.

**P.H.A.D**: Project administration. **M.P.P**: Conceptualization, Supervision, writing-original draft, review & editing.

**Data availability** The necessary raw and processed data to replicate our study's results are sourced from Ref. [1]. Our curated data files can be downloaded from the following weblink: https://github.com/sree369nidhi/Machine-Learning-approach-to-predict-phases-and-strength-of-aluminum-containing-refractory-HEA.

# Tables

**Table 1** Comparison of performance of ML models over coefficient of determination ($R^2$), root mean squared error (RMSE), mean absolute error (MAE)

| Model | $R^2$ | RMSE | MAE | Time (s) |
|---|---|---|---|---|
| CatBoost Regressor | 0.847 | 0.379 | 0.284 | 0.540 |
| Random Forest Regressor | 0.798 | 0.438 | 0.329 | 0.474 |
| Extra Trees Regressor | 0.797 | 0.420 | 0.290 | 0.462 |
| Gradient Boosting Regressor | 0.763 | 0.471 | 0.358 | 0.048 |
| Light Gradient Boosting | 0.653 | 0.570 | 0.438 | 0.106 |
| Linear Regression | 0.650 | 0.568 | 0.438 | 0.574 |
| Bayesian Ridge | 0.642 | 0.574 | 0.441 | 0.018 |
| Ridge Regression | 0.640 | 0.576 | 0.443 | 0.014 |
| AdaBoost Regressor | 0.639 | 0.578 | 0.472 | 0.082 |
| Decision Tree Regressor | 0.592 | 0.611 | 0.417 | 0.016 |

**Table 2** Lowest RMSE obtained for the hyper parameters: maximum depth and learning rate.

| Maximum depth, Learning rate* | Lowest RMSE for training data | Lowest RMSE for test data |
|---|---|---|
| 5, 0.01 | $7 \times 10^{-2}$ | $8 \times 10^{-1}$ |
| 6, 0.29 | $3 \times 10^{-2}$ | $3 \times 10^{-1}$ |
| 7, 0.5 | $2 \times 10^{-1}$ | $5 \times 10^{-1}$ |

*The learning rate was varied from 0.01 to 0.9 at each maximum depth. The iterations were stopped if there was no change in RMSE for 100 iterations.

**Table 3** 5-fold cross validation for the CatBoost regressor model

| Folds | $R^2$ |
|---|---|
| 1 | 0.9941 |
| 2 | 0.9868 |
| 3 | 0.9573 |
| 4 | 0.9721 |
| 5 | 0.9816 |
| **Average** | 0.9783 |

**Table 4** New compositions of RHEAs, which are not in the database, to test CatBoost models

| Sr. No. | Alloy | Al | Nb | Ti | Zr | V | Mo | Ta | Phase |
|---|---|---|---|---|---|---|---|---|---|
| 1 | Al0NbTiMoV | 0 | 25 | 25 | 0 | 25 | 25 | 0 | BCC |
| 2 | Al0.25NbTiMoV | 5.4 | 25.5 | 22 | 0 | 22.1 | 25 | 0 | BCC |
| 3 | Al0.5NbTiMoV | 9.8 | 24.3 | 19.9 | 0 | 20.4 | 25.6 | 0 | BCC |
| 4 | Al0.75NbTiMoV | 14.2 | 22.5 | 20.7 | 0 | 20.5 | 22.2 | 0 | BCC |
| 5 | AlNbTiMoV | 17.6 | 21.9 | 16.9 | 0 | 19 | 24.6 | 0 | BCC |
| 6 | Al1.5NbTiMoV | 27.7 | 18.2 | 16 | 0 | 17.8 | 20.4 | 0 | BCC |
| 7 | AlMo0.5NbTa0.5TiZr | 19.9 | 22.9 | 18.5 | 16.8 | 0 | 11.3 | 10.6 | BCC + BCC |
| 8 | AlMo0.5NbTa0.5TiZr | 5.9 | 22.9 | 27.7 | 26.7 | 0 | 0 | 16.8 | BCC + BCC |
| 9 | Al0.3NbTa0.8Ti1.4V0.2Zr1.3 | 5.1 | 22.2 | 27.7 | 27.8 | 2.5 | 0 | 12.6 | BCC |
| 10 | Al0.5NbTa0.8Ti1.5V0.2Zr | 11.4 | 22 | 28.2 | 20.6 | 4.7 | 0 | 13.1 | BCC + BCC |

\* PC=Processing condition, A= Annealing, AC=Air cooling. HIP=Hot Isostatic pressing. Sr. No 1-6 are from Ref.[24] and 7-10 are from Ref.[16]

**Table 5** Performance of ML classifier models

| Model | Accuracy | Time (s) |
|---|---|---|
| CatBoost | 0.885 | 1.164 |
| Decision Tree | 0.855 | 0.028 |
| Random Forest | 0.832 | 0.604 |
| Extra Trees | 0.821 | 0.528 |
| Logistic Regression | 0.820 | 1.074 |
| Gradient Boosting Classifier | 0.806 | 0.566 |
| Light Gradient Boosting | 0.775 | 0.172 |
| Naïve Bayes | 0.710 | 0.020 |
| K Neighbors Classifier | 0.676 | 0.140 |
| Ridge Classifier | 0.644 | 0.034 |
| AdaBoost Classifier | 0.501 | 0.124 |

**Table 6** Hyperparameter optimization for CatBoost classifier

| Maximum depth, Learning rate* | Lowest MultiClass loss for training data | Lowest MultiClass loss for test data |
|---|---|---|
| 4, 0.07 | $7 \times 10^{-1}$ | $11 \times 10^{-1}$ |
| 6, 0.10 | $9 \times 10^{-3}$ | $1 \times 10^{-2}$ |
| 8, 0.03 | $11 \times 10^{-3}$ | $9 \times 10^{-2}$ |

*The learning rate was varied from 0.01 to 0.9 at each maximum depth

**Table 7**  5-fold Crossvalidation for CatBoost classifier

| Folds | $R^2$ |
|:-----:|:------:|
| 1 | 0.9444 |
| 2 | 0.9988 |
| 3 | 0.9587 |
| 4 | 0.9863 |
| 5 | 0.9921 |
| **Average** | 0.9760 |

# Graphical Abstract

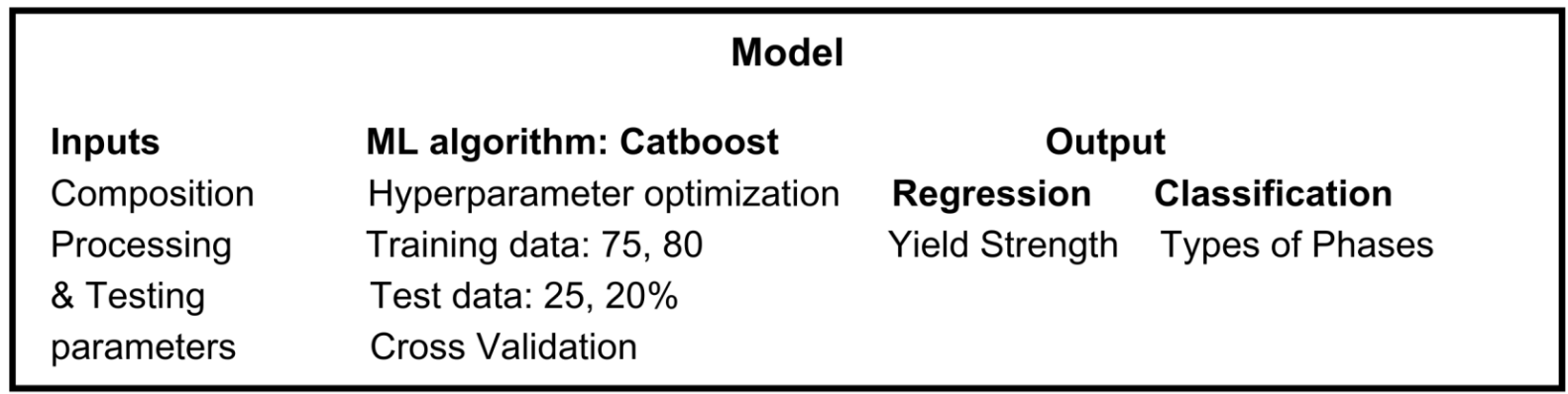
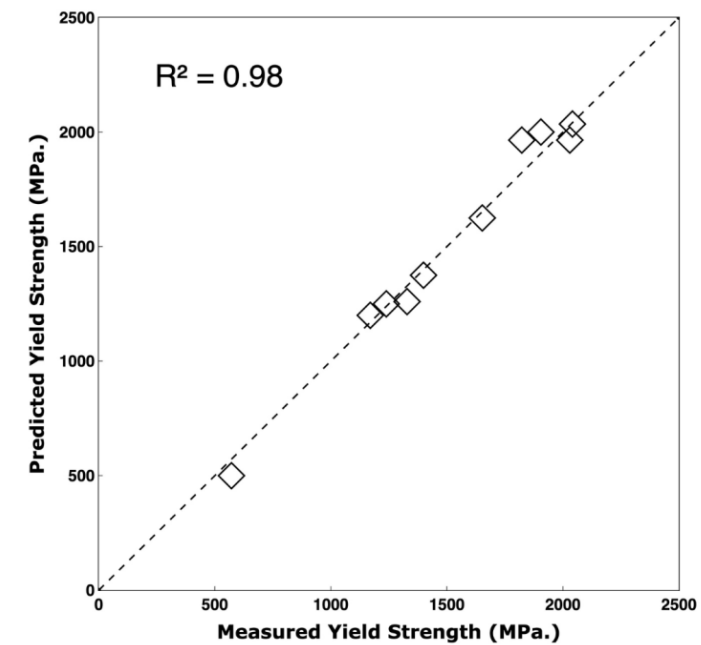
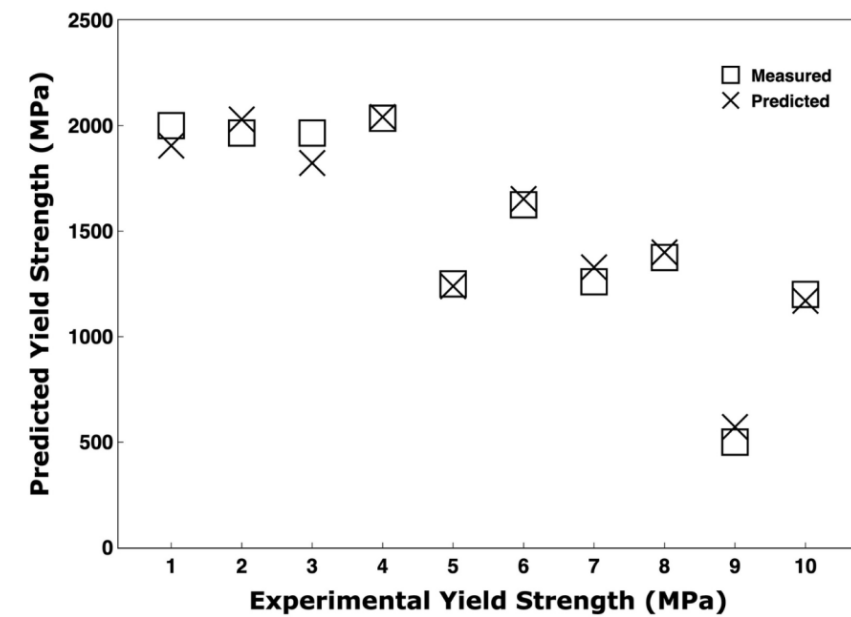
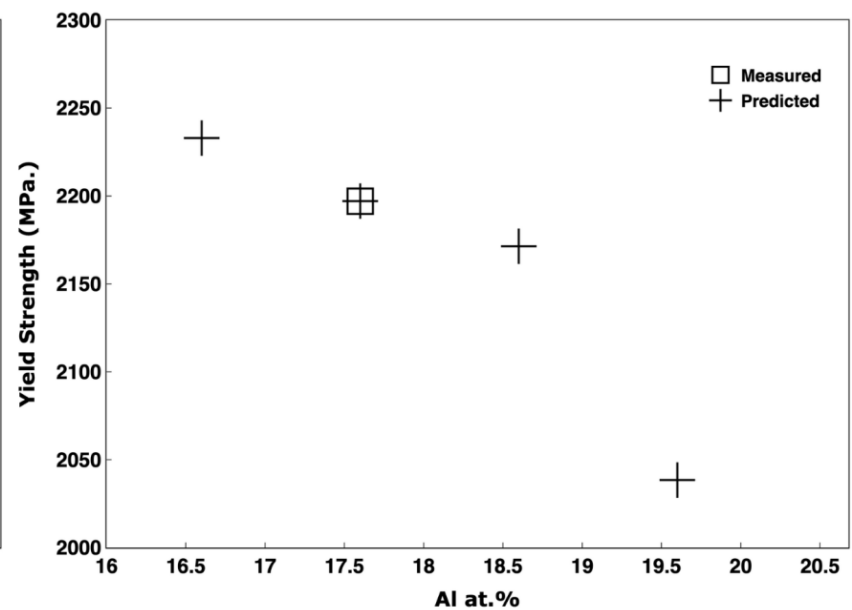
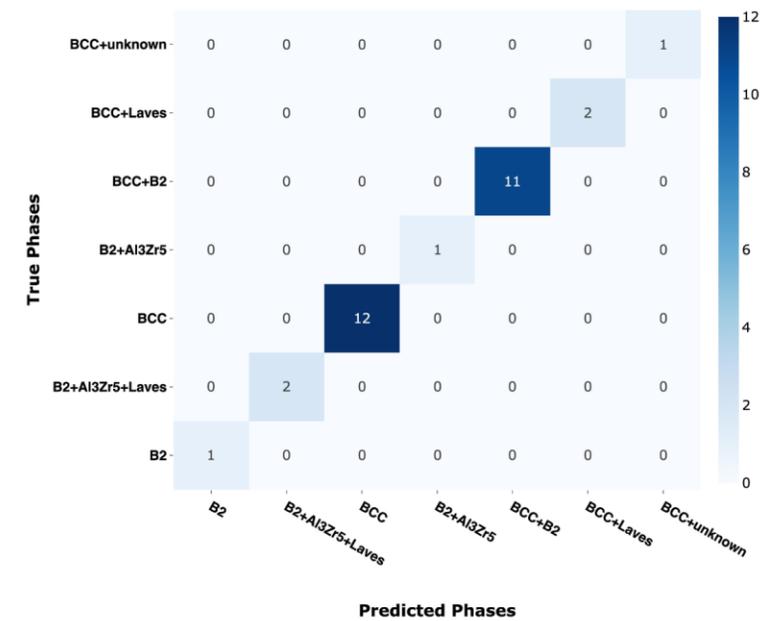

**Highlights**

1. CatBoost model has the best performance among various ML models
2. The model is cross-validated and predicts yield strength accurately ($R^2=0.98$)
3. The effect of aluminum on yield strength is predicted RHEA reasonably
4. Predicts strength for new compositions, that are not in the database
5. The classifier predicts phases formed accurately.

**Figure 1** The range of elemental composition of the alloys in the database. The dashed lines indicate the elemental composition range common to all alloys.

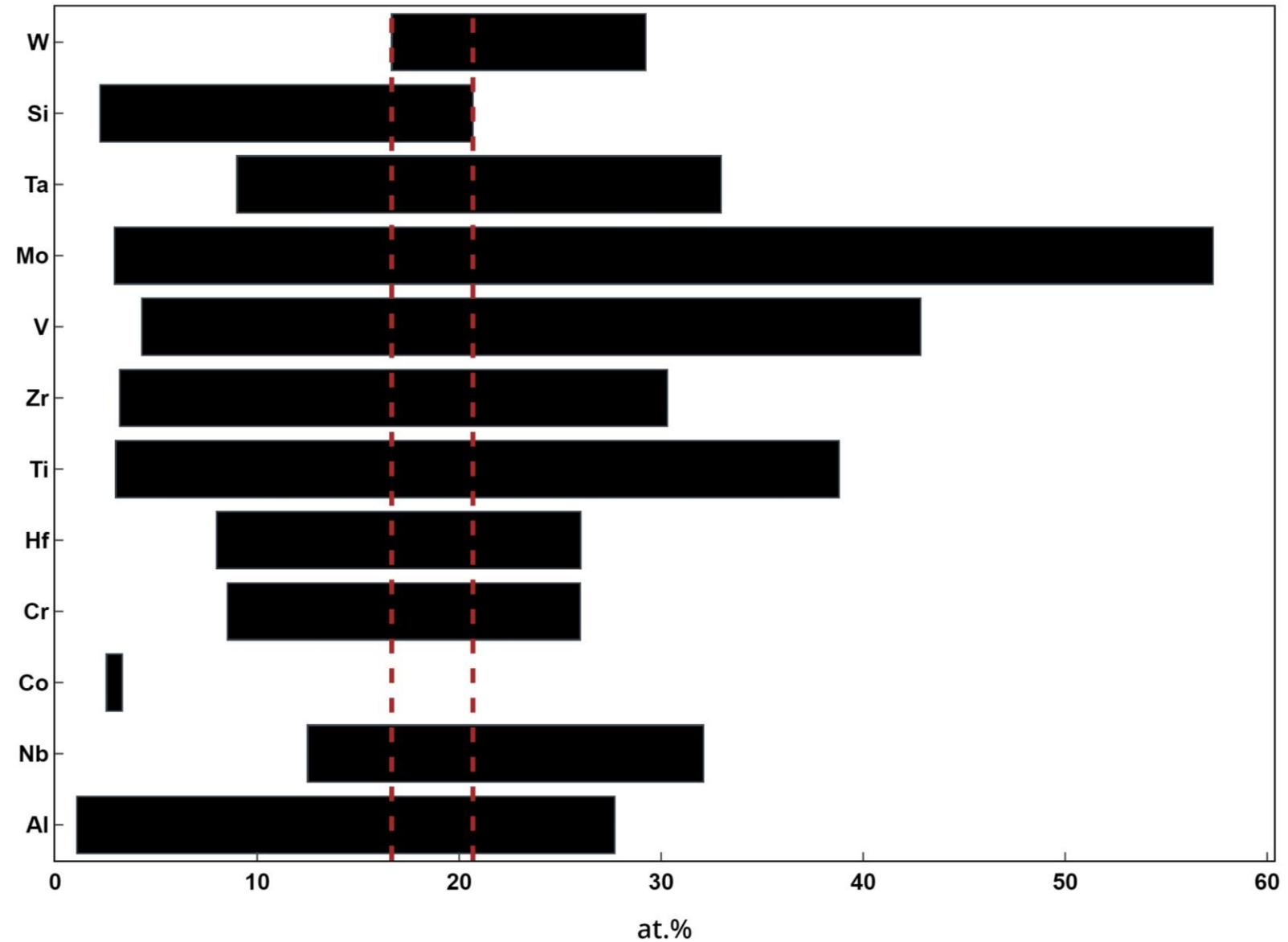

**Figure 2** a) Color map indicating the coefficient of correlation between variables,

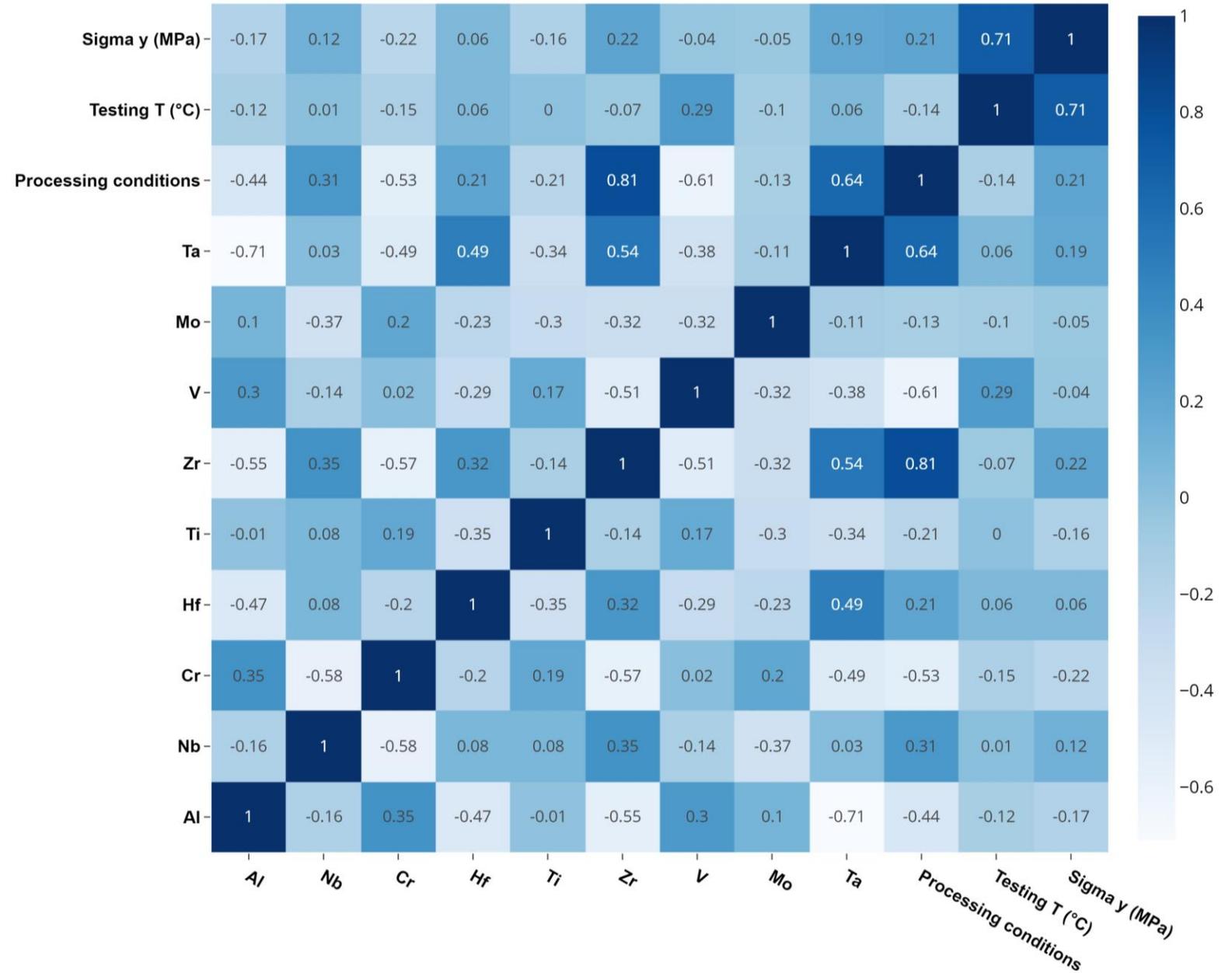

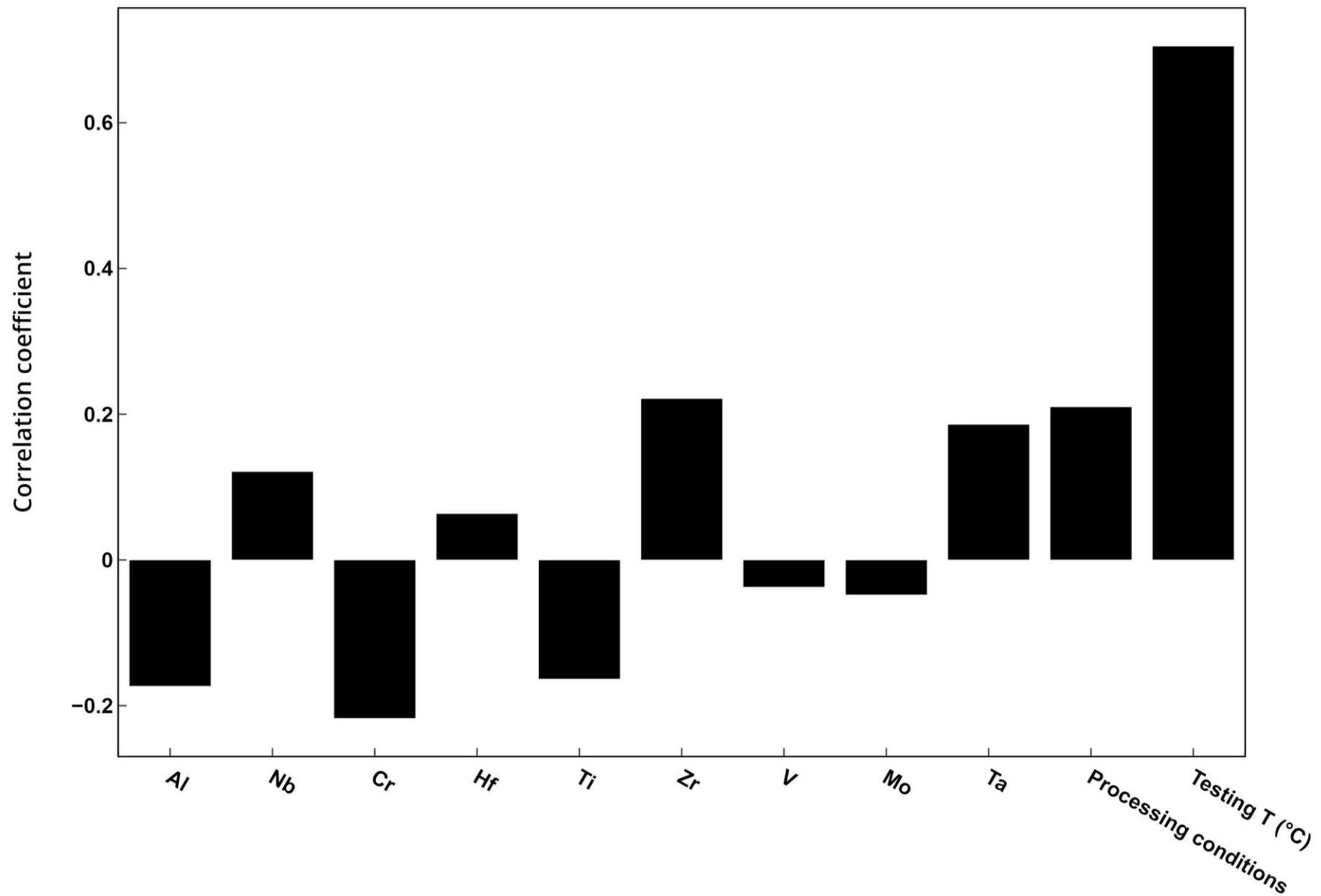

**Figure 2** b) Correlation coefficient between input variables and yield strength.

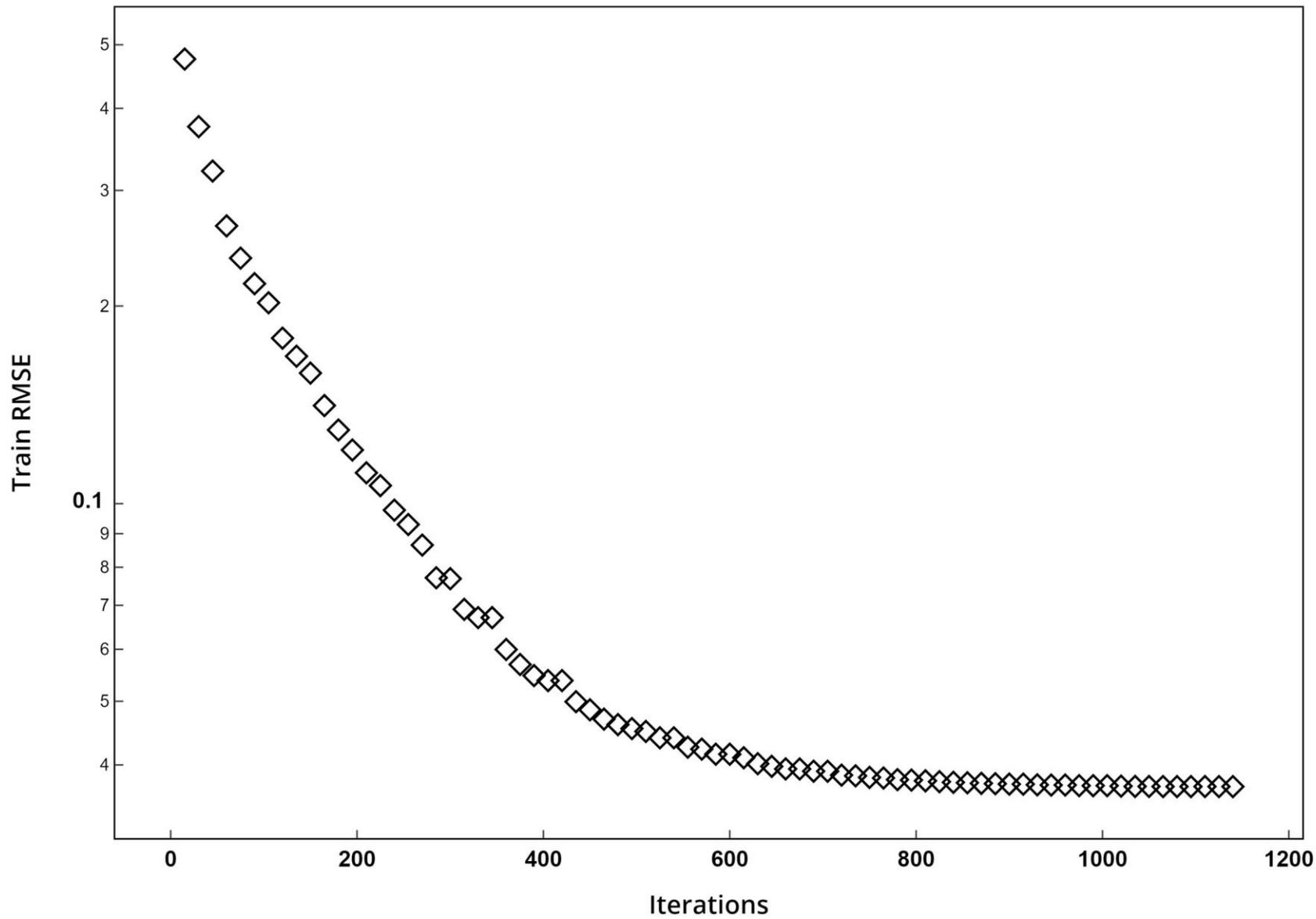

**Figure 3** a) The decrease in RMSE during training for the CatBoost model with increasing iteration number

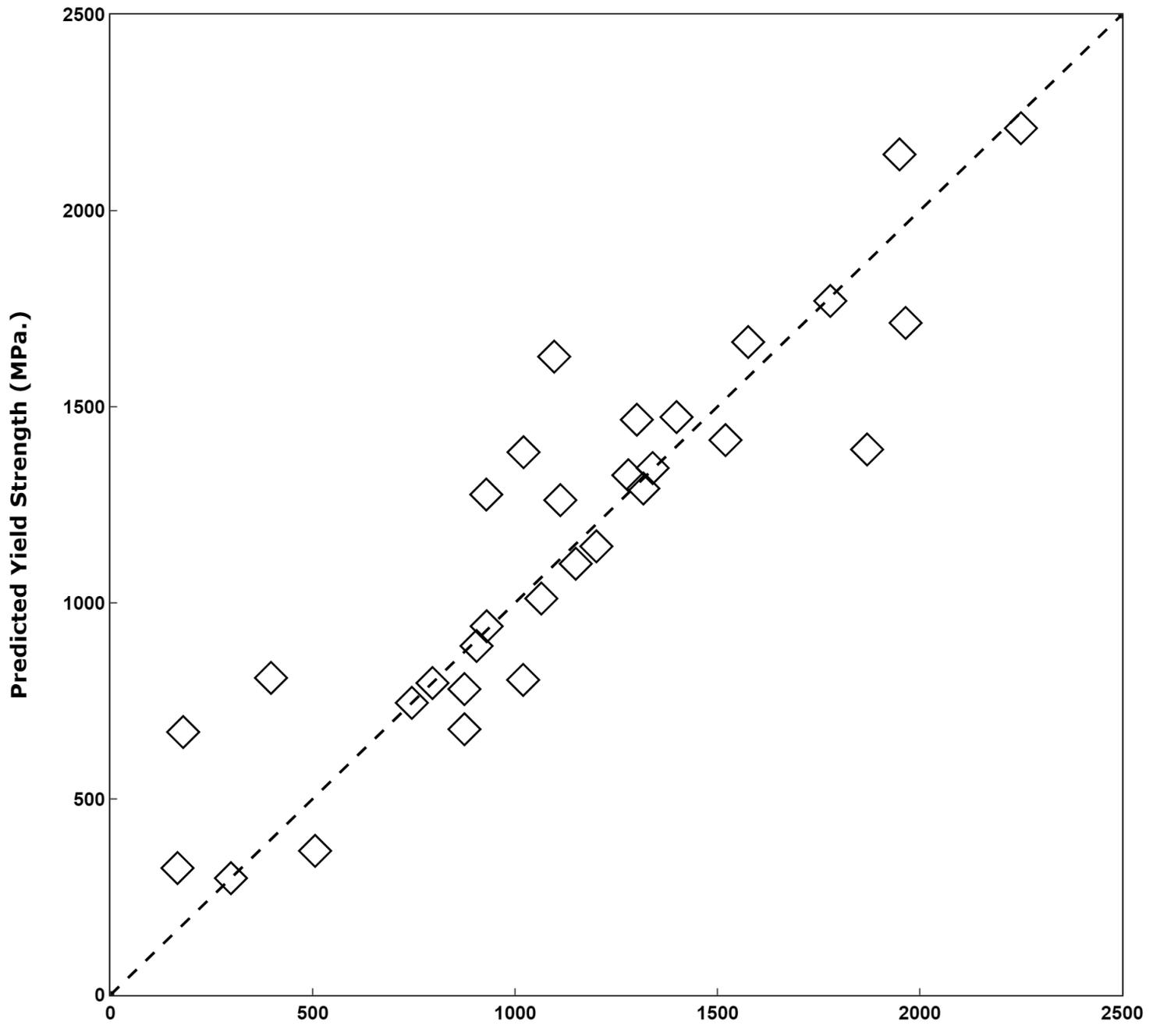

**Figure 3** b) Comparison of predicted and measured yield strength for the test data.

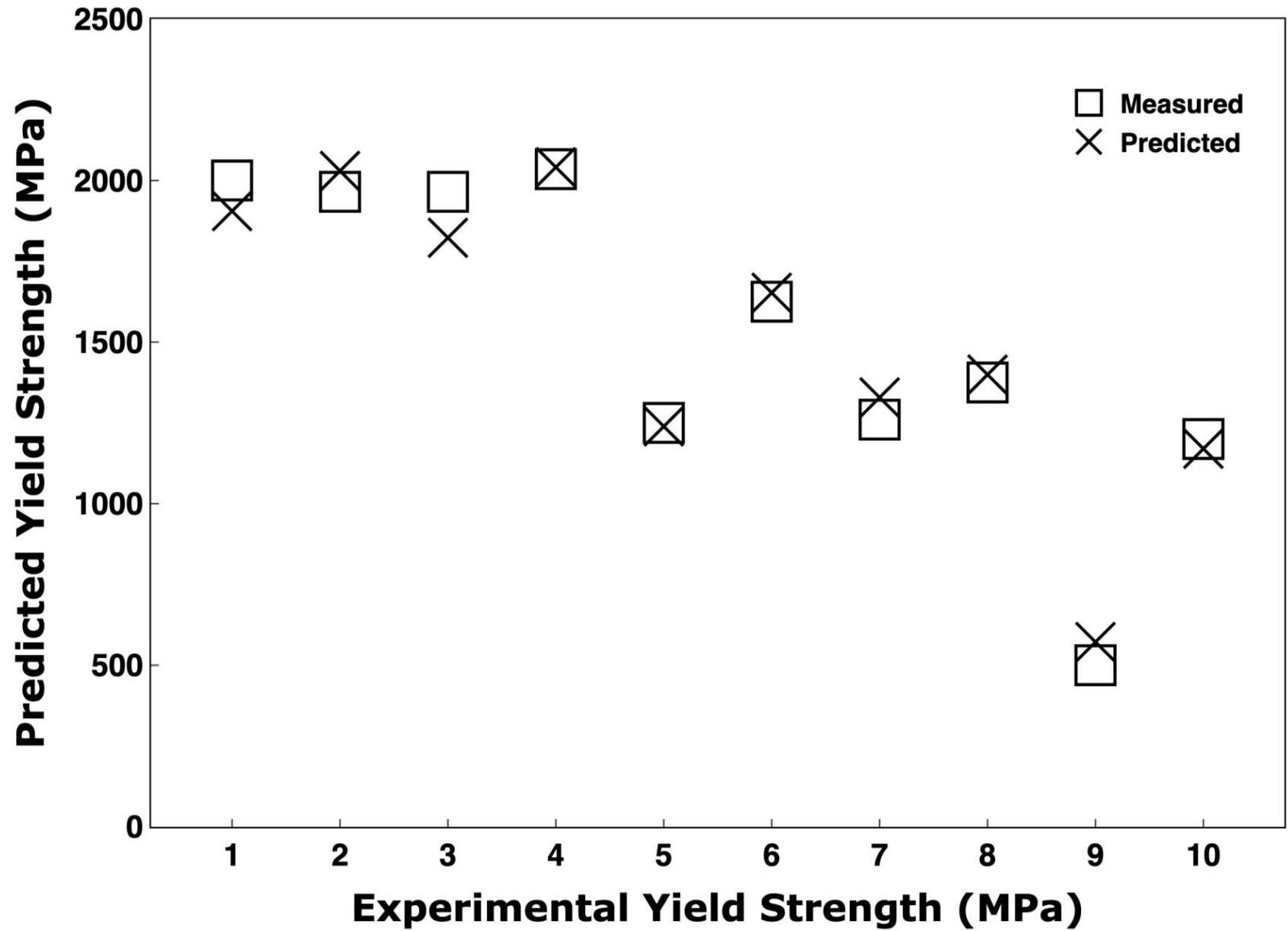

**Figure 4** Comparison of predicted and measured yield strength for alloy composition numbers 1-6 and 7-10 in Table 4, respectively.

**Figure 5** Effect of aluminum content on yield strength of $Al_xMo0.5NbTa0.5TiZr$

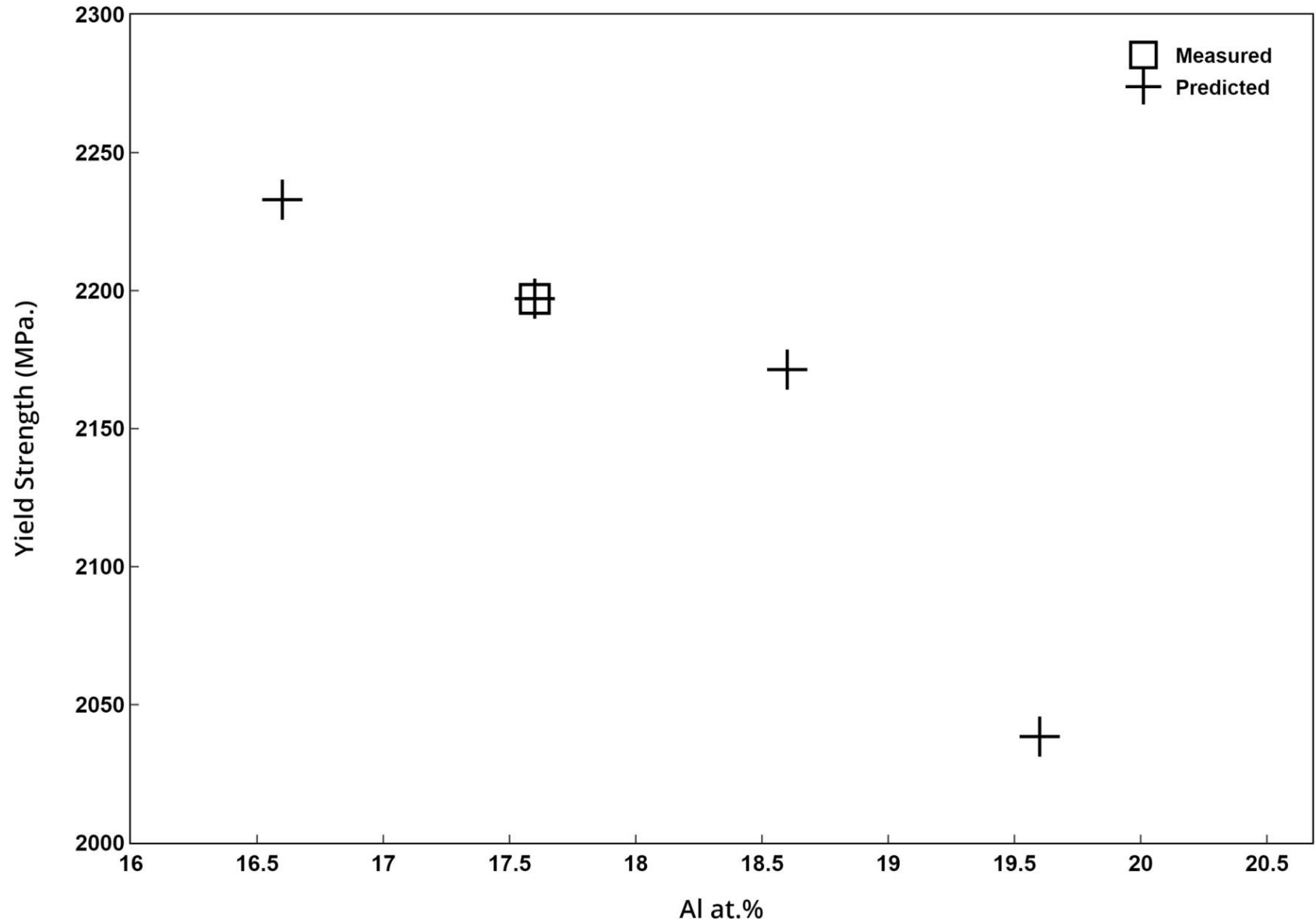

**Figure 6** Comparison of phase predictions with experimentally observed phases for the alloys in the test data presented in the form of a confusion matrix.

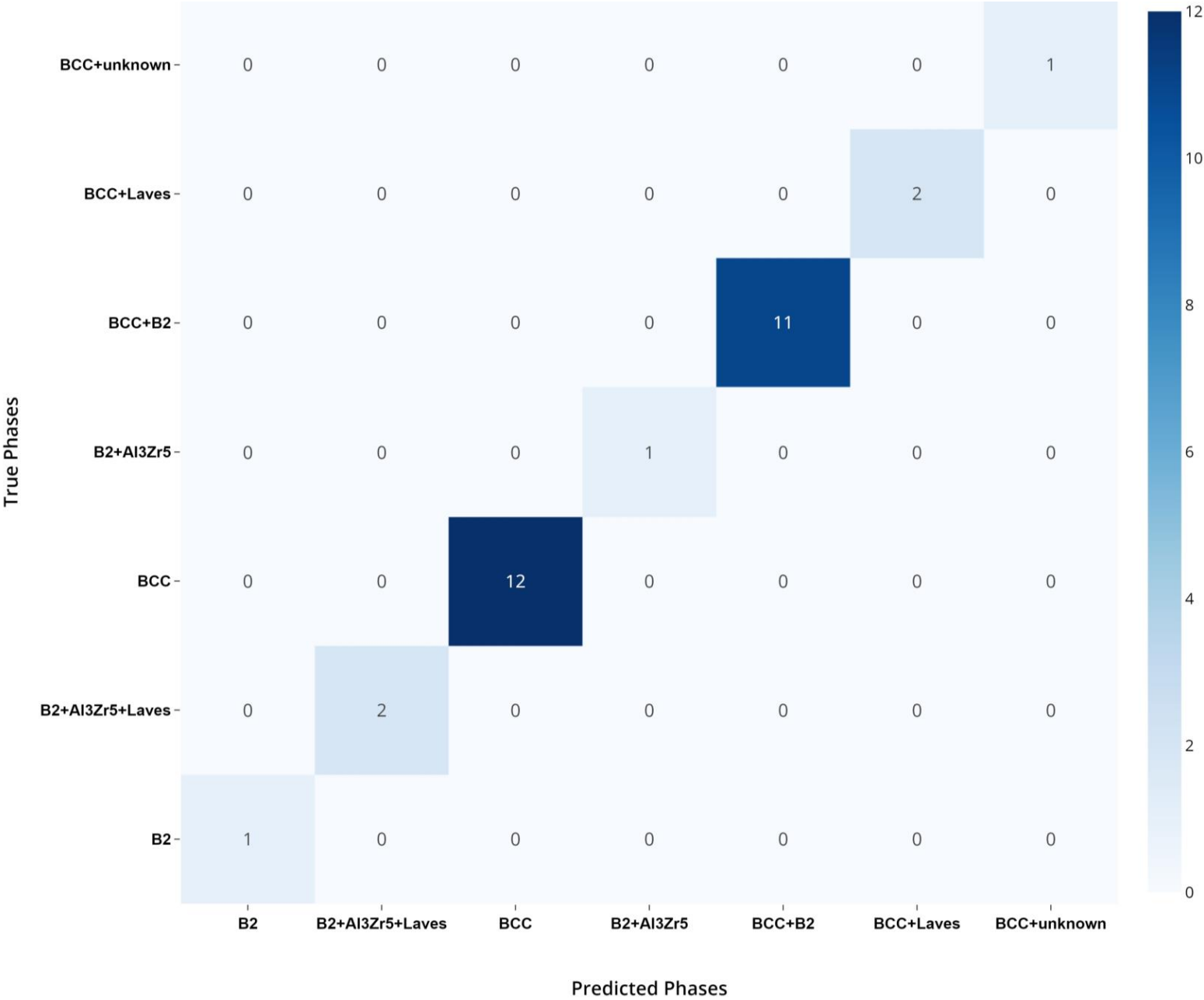

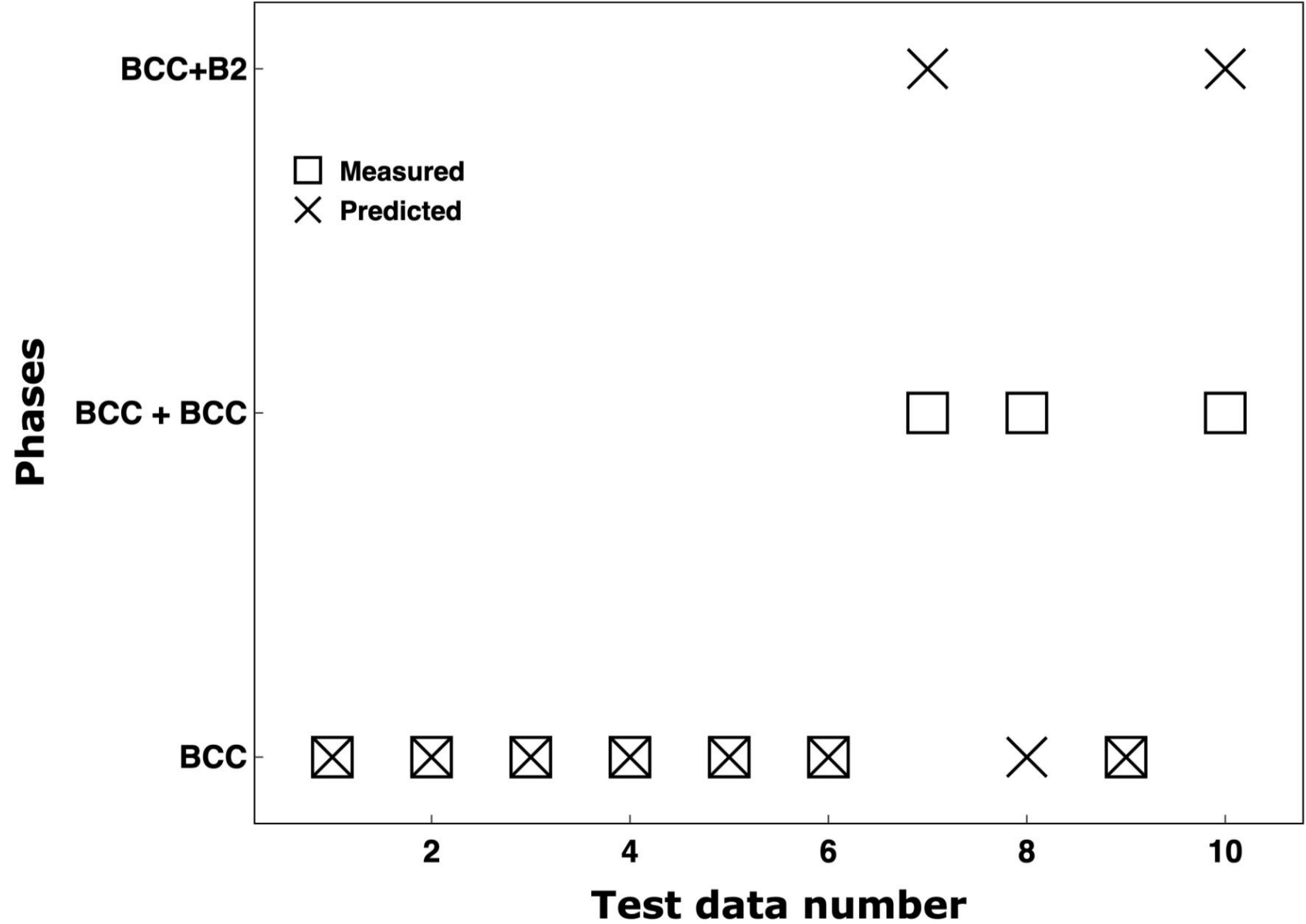

**Figure 7** Phase prediction by CatBoost classifier for new alloy compositions. The numbers on the x-axis indicate the serial numbers of alloy compositions that are given in Table 4